\documentclass[10pt,a4paper,final]{iopart}
\usepackage{iopams}
\usepackage{graphicx}
\usepackage[breaklinks=true,colorlinks=true,linkcolor=blue,urlcolor=blue,citecolor=blue]{hyperref}

\begin{document}

\title[]{Progress on accurate measurement of the Planck constant: watt balance and counting atoms}
\author{Li Shi-Song$^{a)b)\dag}$, Zhang Zhong-Hua$^{b)\dag}$, Zhao Wei$^{a)}$, Li Zheng-Kun$^{b)}$, Huang Song-Ling$^{a)}$}
\address{$^a)$Department of Electrical Engineering, Tsinghua University, Beijing 100084, China}
\address{$^b)$National Institute of Metrology, Beijing 100013, China}
\ead{leeshisong@sina.com; zzh@nim.ac.cn}

\begin{abstract}
The Planck constant $h$ is one of the most significant constants in quantum physics. Recently, the precision measurement of the numeral value of $h$ has been a hot issue due to its important role in establishment for both a new SI and a revised fundamental physical constant system. Up to date, two approaches, the watt balance and counting atoms, have been employed to determine the Planck constant at a level of several parts in $10^8$. In this paper, the principle and progress on precision measurement of the Planck constant using watt balance and counting atoms at national metrology institutes are reviewed. Further improvements for the Planck constant determination and possible developments of a revised physical constant system in future are discussed.
\end{abstract}

\vspace{2pc}
\submitto{\bf Chinese Physics B}
\section{Introduction}  
The Planck constant, $h$, is a fundamental physical constant in the beginning defined to describe the proportionality between the energy $E$ of a charged atomic oscillator in the black body and its frequency $f$, which can be expressed as $E=hf$ \cite{planck1901ueber}. Later in modern physics, some other important roles of the Planck constant were found in quantum mechanics, e.g., $h$ is a key linkage in the quantum Hall effect (QHE) \cite{klitzing1980new}, the Josephson effect (JE) \cite{josephson1962possible}, etc. Based on its fundamental status in quantum physics, experiments obtaining the numeral value of the Planck constant within the International System of Units (SI) have been carried out more than one hundred years \cite{steiner2013history}.

Recently, the accurate measurement of the Planck constant has been focused and will play a most important role in establishment of the new SI \cite{mills2006redefinition,Davis2014}. As the fast development of quantum physics in last century, many macroscopic quantum phenomenons have been employed and successfully built into quantum standards for maintaining and representing SI units \cite{Giovannetti2006,Xiang2013}. The atomic clock is applied for high precision maintenance of the SI base unit, the second, whose measurement accuracy is at a level of at least of several parts in $10^{16}$ \cite{hinkley2013atomic}, leading wide applications in human daily life, e.g., GPS positioning \cite{Chao2010}, meteorology \cite{Bi2013}, navigation \cite{ZhangHL2012}, etc. New electrical standards, the quantum Hall resistance (QHR) standard \cite{novoselov2006unconventional} and the Josephson voltage standard (JVS) \cite{Benz1996}, replaced the conventional artifact standards, i.e., a certain type of stable resistor and voltage cell, making the metrological accuracy for electrical quantities improved at least two magnitudes \cite{Taylor1989}. However, the kilogram, unit of mass, has been the last SI base unit that is kept by artifact, the International Prototype of Kilogram (IPK). The known problem for the artifact standard is its possible drift over time. For IPK, it is believed a change of at least 50$\mu$g happened during one century from 1889 to 1989 \cite{Girard1994}. In order to eliminate the last artifact SI base unit, possible quantum realizations of mass have been tried. After more than thirty years of efforts, the redefinition of the kilogram by fixing the numeral value of the Planck constant $h$ has been widely accepted in both metrology and physical community \cite{Mills2011A, Bord2005A}.

The new definition of the kilogram would be adopted as early as in 2018 \cite{Milton2014}, and it defines the kilogram as "the kilogram, kg, is the unit of mass; its magnitude is set by fixing the numerical value of the Planck constant to be equal to exactly $6.62606X\times10^{-34}$Js when it is expressed in the unit s$^{-1}\cdot$m$^2\cdot$kg, which is equal to J$\cdot$s" \cite{mills2006redefinition}, where $X$ is the last few digits of $h$. By the recommendation of International Committee for Weights and Measures (CIPM), the above definition can be realized only when the Planck constant is measured with a relative uncertainty of several parts in $10^8$.

In the meantime, an accurate measurement of the Planck constant would lead to a more precision physical constant system. It can be seen from the latest evaluation report from the Committee on Data for Science and Technology (CODATA) \cite{CODATA2010}, more than 99\% uncertainty of some other physical constants, e.g., the Avogadro constant $N_A$, the electron charge $e$, the rest mass of the electron $m_e$, the Bohr magneton $\mu_B$ and the nuclear magneton $\mu_N$, is contributed by the uncertainty component of the Planck constant. Therefore, improving the measurement accuracy of the Planck constant can synchronously reduce the uncertainties of these related constants \cite{shisong2012}.

The accurate measurement of the Planck constant was selected as one of the most difficult scientific problems in the worldwide research in 2012 \cite{Jones2012}, since any of such approaches should combine the most precision techniques in electrical, mechanical, optical and chemical metrology. A lot of early approaches were tried in history, e.g., the ampere balance \cite{Vigoureux1965}, the voltage balance \cite{Sienknecht1986}, the Faraday constant method \cite{Bower1980}, etc; however, they were later proved to be  metrologically limited and not persuaded further.
Up to today, two ongoing approaches, watt balance \cite{Kibble1976,Eichenberger2009,Stock2013R} and counting atoms \cite{Andreas2011PRL,Bettin2013R,Mana2012R} (also known as the X-ray crystal density method or the Avogadro project), have been proved as the most precise experiments that can measure the Planck constant at a level of several parts in $10^8$.
In this paper, both progresses of the watt balance experiment and the Avogadro project are reviewed. Suggestions for improving the experiments are discussed. By mapping the new SI to physics laws, possible developments of the physical constant system in future are predicted.
\section{Approach principles}
\subsection{The watt balance principle}
The origin of the watt balance was based on a conventional ampere balance experiment that was designed for absolute measurement of the base unit, the ampere \cite{Vigoureux1965}. The ampere balance was realized by balancing an electrical force and the gravity of a test mass $m$, written as
\begin{equation}
\frac{\partial M}{\partial z}I_1I_2=mg,
\end{equation}
where $I_1$ and $I_2$ are DC currents through the primary and secondary coils; $M$ is the mutual inductance between two coils; $g$ is the acceleration due to the gravitation. The realization of the ampere balance experiment was simple; however, the measurement accuracy was limited by the largest uncertainty component arising from the geometrical factor $\partial M/\partial z$. Since no direct measurement techniques were available for precision determination of the geometrical factor at that time, $\partial M/\partial z$ was determined by dimensional calculations according to Maxwell electromagnetic equations. Besides, the coil was a typical 3-dimension system and difficult to achieve a good perfection when multi-layers were used. In reality, the coil used in ampere balance was made only a few layers to ensure the geometrical uniformity for the calculation, yielding a few grams of magnetic force. As a result, it was very difficult for the ampere balance to measure a mass at kilogram level and the typical uncertainty achieved was about one part in $10^5$.

In 1975, B. P. Kibble at National Physical Laboratory (NPL, UK) proposed a watt balance idea by dividing the experiment into two separated modes \cite{Kibble1976}: the weighing mode and the velocity mode. The weighing mode of a watt balance is operated similar to the ampere balance with much stronger magnetic field $B$, expressed as
\begin{equation}
BLI=mg,
\label{eq.weighing}
\end{equation}
where $L$ is the coil wire length; $I$ is the current through the coil. In the velocity mode, the coil is moved with a velocity $v$ in the magnetic field, yielding an induced voltage $\varepsilon$ as
\begin{equation}
BLv=\varepsilon.
\label{eq.velocity}
\end{equation}
By a combination of equation (\ref{eq.weighing}) and equation (\ref{eq.velocity}), the geometrical factor $BL$ is eliminated, and a virtual power comparison equation, relating the electrical power to mechanical power, is obtained as
\begin{equation}
mgv=\varepsilon I.
\label{eq.wattbalance}
\end{equation}
In equation (\ref{eq.wattbalance}), the induced voltage $\varepsilon$ is measured by a JVS linked to the Josephson effect as
\begin{equation}
\varepsilon=\frac{f_1h}{2e},
\label{eq.voltage}
\end{equation}
where $f_1$ is a known frequency; $e$ is the electron charge. The current $I$ is measured by the Josephson effect in conjunction with the quantum Hall effect as
\begin{equation}
I=\frac{U}{R}=\frac{f_2h}{2e}\frac{ne^2}{h}=\frac{f_2ne}{2},
\label{eq.current}
\end{equation}
where $U$ is the voltage drop on a resistor $R$ in series with the coil; $f_2$ is a known frequency and $n$ is a integer number. Equations (\ref{eq.wattbalance}), (\ref{eq.voltage}) and (\ref{eq.current}) lead to a relation for determining the Planck constant $h$ in SI units as
\begin{equation}
h=\frac{4mgv}{f_1f_2n}.
\label{eq.h}
\end{equation}

All quantities on the right side of equation (\ref{eq.h}) can be precisely determined with a relative uncertainty lower than one part in $10^8$: $m$ is traced to IPK by mass comparators at a level of several parts in $10^9$ \cite{Quinn1992}; $g$ is directly measured by commercial gravimeters at $10^{-9}$ level \cite{Poli2011}; $v$ is determined by measuring the coil displacement using interferometers at sub-nanometer level in a total range of tens of millimeter \cite{Williams1998}; uncertainties of $f_1$, $f_2$, and $n$ are much smaller, which can be neglectful compared to $1\times10^{-9}$ \cite{hinkley2013atomic}. As a conclusion, on a well aligned watt balance platform, the Planck constant is expected to be determined with a relative uncertainty less than $2\times10^{-8}$.

\subsection{The principle of counting atoms}
\label{NA}
The counting atoms \cite{Andreas2011PRL}, also known as the X-ray crystal density (XRCD) method or the Avogadro project, is an indirect approach for precision measurement of the Planck constant. In the approach, the Avogadro constant $N_A$ is first measured by counting Si atoms in a purified silicon sphere, then the Planck constant is determined based on the product of $N_Ah$, whose value can be determined much more precise than the goal of the Planck constant determination \cite{CODATA2010}. The $N_Ah$ product meets the following relationship as
\begin{equation}
N_Ah=\frac{c_0\alpha^2A_r(e)M_u}{2R_\infty}.
\label{eq.Nah}
\end{equation}
In equation (\ref{eq.Nah}), $A_r(e)$ is the electron relative atomic mass with a relative uncertainty of $u=4\times10^{-10}$; $M_u$ is the molar mass constant (defined as fixed constant); $R_{\infty}$ is the Rydberg constant, $u=5\times10^{-12}$; $c_0$ is the speed of light in vacuum (defined as fixed constant); $\alpha$ is the fine structure constant, $u=3.2\times10^{-10}$. According to the latest CODATA evaluation \cite{CODATA2010}, the combined measurement uncertainty for $N_Ah$ is $7\times10^{-10}$, which is more than 60 times smaller than the uncertainty of the Planck constant $u=4.4\times10^{-8}$.

\begin{figure}[h]
\center
\includegraphics[width=1.6in]{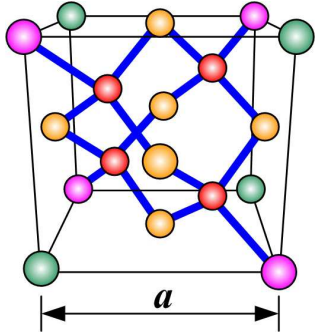}
\caption{\small The lattice structure of the $^{28}$Si crystal. 18 $^{28}$Si atoms are shown in the structure: 4 atoms (red) are inside the cube; 6 atoms (yellow) are at the centres of 6 surfaces; the other 8 atoms (pink and green) are at vertices of the cube. The average number of atoms in each lattice $N=4+6/2+8/8=8$.}
\label{Fig.1}
\end{figure}

The history and early efforts towards precision determination of the Avogadro constant by counting atoms can be found in \cite{Becker2001}. The silicon sphere was introduced in the Avogadro project for the smaller thermal expansion coefficient and more stable surface oxide layer compared to that of other semiconductor materials, e.g., Ge. During the approach, the natural Si crystal material was used to make the silicon spheres in the beginning; however, some metrological limits were found due to its impurity \cite{Becker2003Determination}. Later the enriched $^{28}$Si single-crystal material was applied to obtain a higher degree of perfection \cite{Becker2006L}. The lattice structure of the $^{28}$Si crystal is shown in figure \ref{Fig.1}. Each lattice is a cube with a lattice parameter $a$, containing 8 $^{28}$Si atoms on average. If the mole volume of $^{28}$Si single-crystal $V_m$ is measured, the Avogadro constant then can be written as
\begin{equation}
N_A=\frac{V_m}{V_a}=\frac{8M_0}{\rho a^3},
\label{eq.Nh}
\end{equation}
where $V_a$ is the volume of a signal $^{28}$Si atom, written as $V_a=a^3/8$; $M_0$ is the mole mass of $^{28}$Si crystal and $\rho$ is its density.

The mole mass $M_0$ is determined by measuring the isotope abundances of three isotopes: $^{28}$Si, $^{29}$Si, and $^{30}$Si. In realization, the isotope percentages of the crystal are measured by isotope dilution. $\chi_m$ denotes the $^{m}$Si ($m=28, 29, 30$) isotope percentage of the crystal, and $M_0$ is determined as \cite{Mana2010}
\begin{equation}
M_0=\chi_{28}M(^{28}\mathrm{Si})+\chi_{29}M(^{29}\mathrm{Si})+\chi_{30}M(^{30}\mathrm{Si}).
\label{eq.si}
\end{equation}
The lattice parameter $a$ is calculated from the \{220\} lattice-plane spacing $d_{220}$ of a silicon crystal measured by the X-ray interferometer \cite{Massa2011} using equation
\begin{equation}
a=\sqrt{8}d_{220}.
\end{equation}
The density of silicon sphere $\rho$ is obtained by both mass and volume measurements. The mass of the silicon sphere is traced to the IPK \cite{Picard2011} and the volume is measured by means of optical interferometers \cite{Bartl2011}. In theory, all three quantities $M_0$, $a$, and $\rho$ can be well determined with relative uncertainties of several parts in ${10^9}$. The Planck constant is expected to be indirectly determined with an accuracy of $2\times10^{-8}$ by counting atoms in the silicon sphere.

\section{The watt balance in progress}
Since the proposal in 1975 \cite{Kibble1976}, the watt balance experiment has been widely spread and carried out in many national metrology institutes (NMIs) on the current stage for precision determination of the Planck constant $h$ and in future for maintaining of the SI base unit, the kilogram. These metrology institutes include the NPL, the National Institute of Standards and Technology (NIST, USA), the Swiss Federal Office of Metrology (METAS, Switzerland), the Laboratoire national de m\'{e}trologie et d'essais (LNE, France), the Bureau International des Poids et Mesures (BIPM), the National Institute of Metrology (NIM, China), the National Research Council (NRC, Canada), the Measurement Standards Laboratory (MSL, New Zealand) and the Korea Research Institute of Standards and Science (KRISS, South Korea). Here the progress made at these NMIs are summarized.

\subsection{NPL-NRC watt balance}

The NPL is the institute that proposed and firstly practiced the watt balance idea and its watt balance experiment was launched in 1977. In the following one decade, the first generation of watt balance, named NPL Mark I, was built in air \cite{Kibble1987}. In the approach, a beam balance was employed as the mass comparator in the weighing mode and the vertical motion stage in the velocity mode. A Fe-Co-Ni-Al permanent magnet was applied to generate a 0.7T magnetic field. In the velocity mode, an 8-shape coil with two square segments were moved by 2mm/s to induce a 1V voltage. In the weighing mode, 500g and 1kg masses were added or removed according to the current direction in coil. The first measurement result was reported in \cite{Kibble1987}, $h=6.6260688(37)\times10^{-34}$Js with a relative uncertainty of $5.6\times10^{-7}$. Later an updated value of the Planck constant $h=6.62606821(90)\times10^{-34}$Js with a relative uncertainty of $1.4\times10^{-7}$ was published in \cite{Kibble1990}.

\begin{figure}[h]
\center
\includegraphics[width=3.8in]{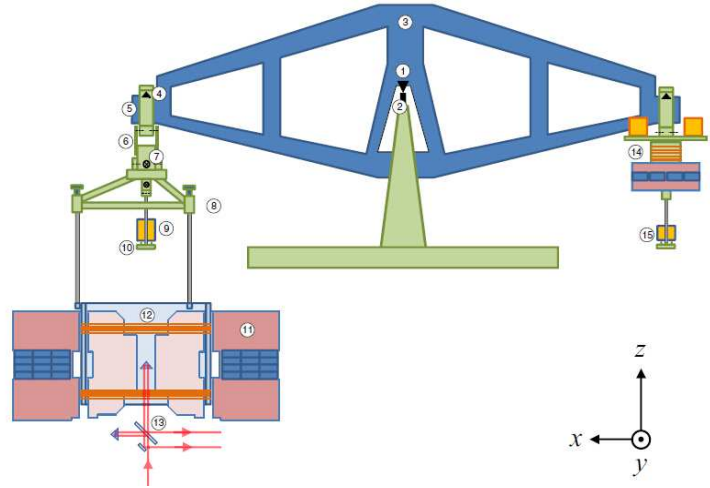}
\caption{\small Schematic of the NPL-NRC watt balance. The numbers denote: 1-central knife-edge, 2-central flat, 3-balance beam, 4-end knife-edge, 5-stirrup,
6-suspension middle section, 7-lower gimbal, 8-suspension lower section, 9-test mass, 10- mass pan, 11-permanent magnet, 12-coil, 13-interferometer, 14-voice coil, 15-tare mass. Reproduced with permission \cite{Sanchez2014}.}
\label{Fig.NPL}
\end{figure}

In 1990, a newly built watt balance, the NPL Mark II (shown in figure \ref{Fig.NPL}), was designed to operate in vacuum. The balance beam was preserved in the Mark II watt balance, whose beam length was 1.2m to reduce the nonlinear motion in the velocity mode \cite{Robinson1997}. Compared to the Mark I watt balance, an important modification for Mark II was that the magnet and the coil were redesigned into horizontal circular for eliminating any linear dependence from coil movement or thermal expansion of the wire \cite{Robinson2007}. The flux of a SmCo magnet were guided to the upper and lower air gaps, generating a magnetic flux density of 0.45T at the coil position. Two opposite series coils were buried in two air gaps. A mirror was attached to the coil former and the Michelson interferometer was used for coil position detection. The voltage drop on a 50$\Omega$ resistor in the weighing mode and the induced voltage in the velocity mode were compared against a programmable Josephson voltage standard (PJVS) while the resistor was calibrated against a quantum Hall resistance standard through a cryogenic current comparator (CCC) bridge. The first published $h=6.62607095(44)\times10^{-34}$Js was produced on NPL Mark II watt balance with a relative standard uncertainty of $6.6\times10^{-8}$ \cite{Robinson2007}. However, after reporting the measurement, an potential mass exchange error source was identified in the weighing mode. As a result, the measurement value was adjusted to $h=6.62607123(133)\times10^{-34}$Js with an expanded uncertainty of $2.0\times10^{-7}$ \cite{Robinson2012}.

Instead of a continuous work on the new generation NPL Mark III watt balance, the NPL decided to shut down the watt balance experiment and sell the Mark II system to NRC. In 2009, the NPL Mark II watt balance was was dismantled and transferred to NRC, Canada. After the reassemble, the watt balance, named NRC watt balance, was independently operated in a newly constructed laboratory. After investigating two effects, the stretching of the coil support flexures under load and the tilting effect on the balance support base due to loading and unloading the mass lift, the NRC watt balance published its first value for the Planck constant $h=6.62607063(43)\times10^{-34}$Js with a relative uncertainty of $6.5\times10^{-8}$ \cite{Steele2012}. In 2014, a updated value of the Planck constant $h=6.62607034(12)\times10^{-34}$Js, which has the smallest uncertainty of $1.9\times10^{-8}$ among the known $h$ determinations, was published in \cite{Sanchez2014}.

\begin{figure}[h]
\center
\includegraphics[width=2.8in]{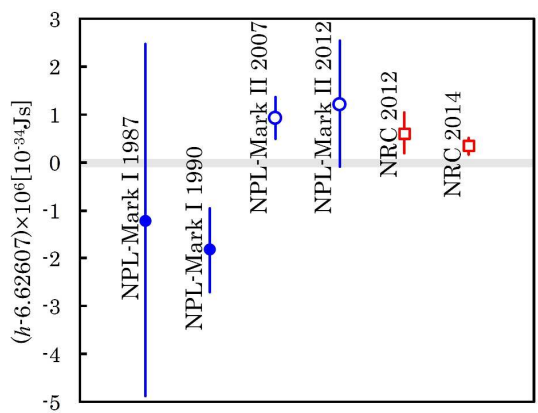}
\caption{\small Results of $h$ determinations by the NPL-NRC watt balance.}
\label{Fig.NPLNRCh}
\end{figure}

As a conclusion, all measurement results of the Planck constant $h$ produced by NPL-NRC watt balances are summarized in figure \ref{Fig.NPLNRCh}. Although the story is a little tortuous, a more than 30 year effort has harvested the word most accurate determination of the Planck constant up to date.

\subsection{NIST watt balance}
The first version of the NIST watt balance, named NIST-1, had transferred from a conventional ampere balance following the watt balance idea since 1980 \cite{Olsen1980}. A rather different design for the NIST watt balance was that an electromagnetic
solenoid was used to generate a radial magnetic field at the coil \cite{Olsen1985}. This realization in which the magnetic field had a $1/r$ contribution along the radial direction was an important conception, since in this case, the experiment would be insensitive to any linear part of both coil horizontal movements and thermal expansion, and only a second-order error should be considered. Note this idea has been widely spread in the watt balance community, including the NPL Mark II magnet design mentioned above. The second difference of the NIST watt balance was that an aluminum wheel was used as their force comparator, which can obviously reduce the nonlinearity than that of the realization by a beam lever. In NIST-1, the solenoid generated a 2.9mT magnetic field with 8A DC current. In the velocity mode, a movable coil was moved with a velocity of 0.667mm/s and a 20mV induced voltage was obtained. In the weighing mode, a 3.33mA DC current was passing through the movable coil and a 15g mass was balanced.
In 1989, NIST-1 produced a value of the Planck constant $6.6260704(88)\times10^{-34}$Js with a relative uncertainty of $1.3\times10^{-6}$ \cite{Cage1989}.

Shortly after publication of NIST-1 result, an updated watt balance apparatus NIST-2 was built, in which a superconducting solenoid was added. A 5A DC current was passing through the new solenoid and a 0.1T magnetic flux intensity was obtained in a 80mm vertical movement interval. A 10mA current was applied in the movable coil in the weighing mode, generating a 5N magnetic force to balance the gravity of a 500g test mass. In the velocity mode, the coil was moved with 2mm/s to generate a 1V induced voltage. After years of finding misalignment errors and apparatus improvements, a measurement of the Planck constant $h=6.62606891(58)\times10^{-34}$Js with a relative uncertainty of $8.7\times10^{-8}$ was published in 1998 \cite{Williams1998}.

\begin{figure}
\center
\includegraphics[width=2.5in]{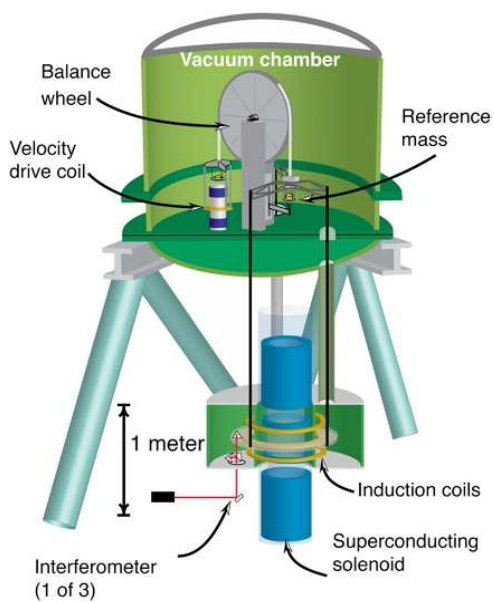}
\caption{\small Apparatus of the NIST-3 watt balance. Reproduced with permission \cite{steiner2013history}. }
\label{Fig.NIST3}
\end{figure}

\begin{figure}
\center
\includegraphics[width=2.8in]{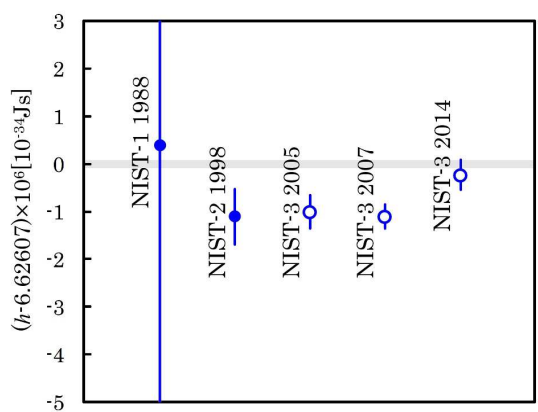}
\caption{\small Results of $h$ determinations by the NIST watt balance.}
\label{Fig.NISTh}
\end{figure}

Since 1999, the NIST-2 watt balance has been updated to NIST-3 (shown in figure \ref{Fig.NIST3}). The main update for NIST-3 was that the watt balance was operated in vacuum instead of in air for NIST-2. In 2005, the numeral value of $h=6.62606901(34)\times10^{-34}$Js with a relative uncertainty of $5.2\times10^{-8}$ was published \cite{Steiner2005}. After some minor improvements and reduction of the type B uncertainty, the Planck constant $h=6.62606891(24)\times10^{-34}$Js with a relative uncertainty of $3.6\times10^{-8}$ was published in 2007 \cite{Steiner2007}. At that point, further improvements to reduce the measurement uncertainty was expected. However, a obvious increase of $9\times10^{-8}$ for the Planck constant determination was observed during March 2010 and May 2010. As part of the reason, the national mass standard K85 in United States was calibrated by BIPM and an increase of $4\times10^8$ was found. Towards a final determination of the Planck constant, measurements of NIST-3 watt balance were made during 2012 to 2013, yielding a number of the Planck constant  $h=6.62606979(30)\times10^{-34}$Js with a relative uncertainty of $4.5\times10^{-8}$ \cite{Schlamminger2014}.

The measurement results of the Planck constant produced from the NIST watt balance have been listed in figure \ref{Fig.NISTh}. No doubt that the NIST watt balance has played an importance role for decades towards the precision measurement of the Planck constant. The NIST watt balance is ongoing currently. A new watt balance, the NIST-4, which employs a permanent magnet structure similar to the magnet design of the BIPM watt balance (section 3.5), is being under construction \cite{Seifert2014}.

\subsection{METAS watt balance}
\begin{figure}
\center
\includegraphics[width=2.8in]{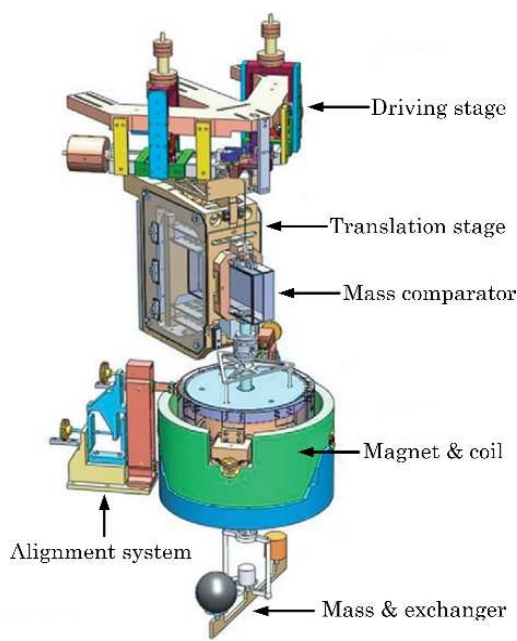}
\caption{\small Design of the METAS watt balance Mark II. Reproduced with permission \cite{Baumann2013}.}
\label{Fig.METAS}
\end{figure}

The METAS watt balance started in 1997. Their first generation watt balance, the METAS Mark I, was aiming to determine the Planck constant by a compact apparatus design \cite{Beer1999, Beer2001, Beer2003}. In realization,
a duel seesaw mechanism was used to move the coil in a 46mm vertical interval. This design had advantages of avoiding both the nonlinear motion of a beam lever and the hysteresis of a knife-edge. Different from either the beam balance in NPL watt balance or the wheel design in NIST watt balance, a commercial weighing cell was employed in METAS Mark I as the force comparator.
As a rather large mechanical load was added on the mass comparator, the test mass used in METAS Mark I watt balance was limited to 100g. The magnetic system was designed similar to that of the NPL Mark I watt balance. The magnetic flux of the SmCo magnet was guided by soft yoke material, generating a 0.56T magnetic field in the air gap. The movable coil was wound in 8-shape with $2\times2000$ turns. A typical magnetic force of 0.5N was generated in the weighing mode when a 3mA DC current was used. In the velocity mode, the coil was moved with 3mm/s, generating a 0.5V induced voltage. In 2011, the METAS Mark I produced a final measurement of the Planck constant $h=6.6260691(20)\times10^{-34}$Js with a relative uncertainty of $2.9\times10^{-7}$ \cite{Eichenberger2011}.

In order to determine the Planck constant with relative uncertainty of several parts in $10^8$, a second generation watt balance, the METAS Mark II (shown in figure \ref{Fig.METAS}), is being under development \cite{Baumann2013}. A driving stage based on a Sarrus linkage has been developed on the new platform. Experimental test showed that deviations from verticality in both $x$ and $y$ axis were smaller than 2$\mu$m. Also, a 13-hinge translation stage was built to guide the mass comparator in the velocity mode. A peak to peak straightness of 190nm in $x$ axis and 40nm in $y$ axis have been achieved in a 35mm vertical movement interval \cite{Cosandier2014}. In METAS Mark II watt balance, a magnet structure introduced by the BIPM watt balance group will be used. To reduce the large thermal effect of the SmCo magnet, a magnetic shunt compensation technique was developed in the new Mark II design. The Fe-Ni alloy shunt would be inserted into the central hole of the magnet, by changing its length, the temperature coefficient can be reduced by a factor of several hundreds. The METAS Mark II watt balance is expected to measure the Planck constant at the $10^{-8}$ level in near future.

\subsection{LNE watt balance}
\begin{figure}
\center
\includegraphics[width=2.4in]{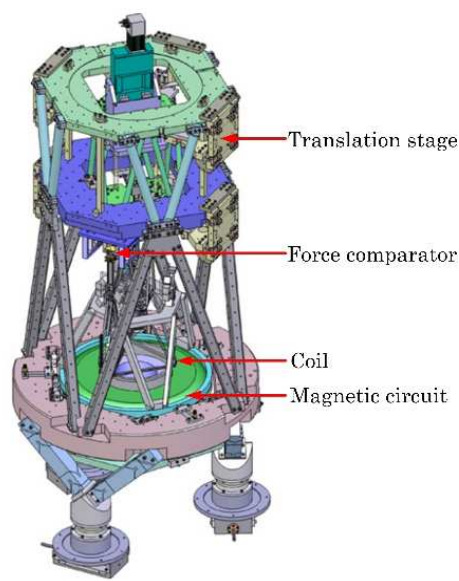}
\caption{\small Design of the LNE watt balance. Reproduced with permission \cite{Villar2011}.}
\label{Fig.LNE}
\end{figure}

The LNE watt balance, shown in figure \ref{Fig.LNE}, began in 2001. One of the main features for the LNE watt balance is that a guiding stage is applied to move both the mass comparator and the coil \cite{Geneves2005}. The idea requires a very strong design of the guiding mechanism with minimizing motions in both $x$ and $y$ directions. In realization, three flexure hinges in every 120$^{\circ}$ were chosen to move the mass comparator and the coil vertically in two planes. Experimental result showed the horizontal motion was less than 0.5$\mu$m along the 75mm travel line in vertical \cite{Villar2011}. The magnet of the LNE watt balance was designed into a one-permanent-magnet, one-coil structure \cite{Gournay2005}. The radial flux density obtained in the air gap was as high as 1T, which has been the maximum magnetic flux density among all existing watt balances. A high magnetic field design can reduce the temperature variation due to power dissipated as well as the nonlinear magnetic errors in the weighing mode. A novel method of velocity control based on the use of a heterodyne Michelson's interferometer, a two-level translation stage, and a homemade high frequency phase-shifting electronic circuit, has been developed and a relative uncertainty of $10^{-9}$ over 60mm was obtained \cite{Topcu2004}. A position sensing system based on the Gaussian beam propagation properties and the spatial modulation was applied in LNE watt balance and a resolution of 25pm/$\sqrt{\mathrm{Hz}}$ has been achieved \cite{Haddad2009}. During the weighing mode, a 5mA DC current though the coil would generate a 5N magnetic force to balance the gravity of a 500g standard mass. In the velocity mode, the velocity was set to 2mm/s, inducing a 1V voltage. Very different from other watt balance projets that using the commercial gravimeter FG5 for determining the local gravitation acceleration, $g$ was measured by a cold atom gravimeter (CAG) in the LNE watt balance \cite{Merlet2008}. Comparison experiment showed that CAG agreed with FG5 in several part in $10^9$ \cite{Merlet2010} with better short term stability \cite{Gillot2014}. The LNE watt balance is under development, e.g., alignment \cite{Thomas2014}, and the Planck constant measurement may report as early as in 2015.

\subsection{BIPM watt balance}
\label{sec.BIPM}
\begin{figure}
\center
\includegraphics[width=2.4in]{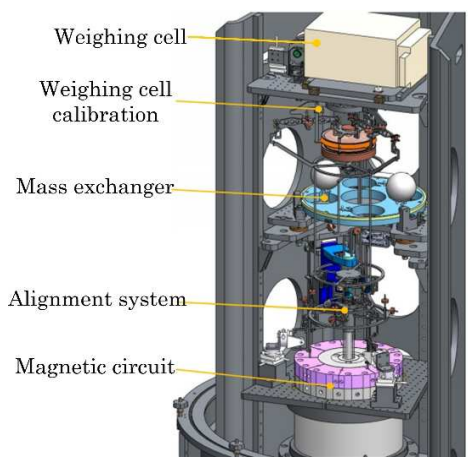}
\caption{\small Design of the BIPM watt balance.}
\label{Fig.BIPM}
\end{figure}

The BIPM watt balance was proposed in 2002 and launched in 2005 \cite{Picard2007}. The schematic of the BIPM watt balance is shown in figure \ref{Fig.BIPM}. Several novel ideas have been practised in BIPM watt balance experiment. Firstly, the simultaneous measurement for the weighing and velocity modes are applied to reduce possible systematic effects that arise from time-varying magnetic flux density and coil misalignment. However, the voltage drop across the coil has an undesired resistive component, which should be eliminated. To remove the resistive voltage drop, several methods including the superconducting coil method \cite{de2014}, bifilar coil method \cite{robinson2012} and data combination method \cite{Fang2013} are being carried out.
The second idea practised in BIPM watt balance is that they use an electrostatic motor to drive a three-arm-lever lifting stage. The electrostatic motor eliminates additional magnetic fields generated from electromagnetic drives; however, the electrostatic force should be considered and well optimized.
A contribution for the BIPM watt balance group is that they have presented a two-permanent-magnet, one-coil constructed magnet, which has a better self magnetic shielding than the one-permanent-magnet, two-coil structure applied in NPL Mark II watt balance. The design later is followed by the METAS Mark II watt balance \cite{Baumann2013}, the NIST-4 watt balance \cite{Seifert2014}, the MSL watt balance \cite{Sutton2014} and the KRISS watt balance \cite{Kim2014}.

In realization, the BIPM watt balance employed a commercial air-compatible 10 kg weighing
cell with 10$\mu$g resolution as the force comparator. A movable coil was set into an air gap with radial magnetic flux density of about 0.5T. In the measurement, a 1mA DC current was injected into the coil, generating a 1N magnetic force to balance a 100g copper mass. In the meanwhile, the coil was moved with a constant velocity of 0.2mm/s, yielding a 0.1V induced voltage. Experimental result showed that the simultaneous weighing and moving approach agreed with the bifilar coil technique in several parts in $10^7$ \cite{Fang2013}. The BIPM watt balance is ongoing. Further improvements including noise reduction, dynamic alignment of the coil, vacuum operation, etc, are being carried out. The Planck constant with an uncertainty of several parts in $10^7$ is expected to be published in 2015.

\subsection{NIM joule balance}
The joule balance experiment at NIM was proposed in 2006. Different from a conventional moving watt balance, the joule balance employs all static phases in the measurement \cite{Zhang2011}. In the weighing mode, the residual magnetic forces $\Delta f$ at different positions are measured. The velocity mode is replaced by the mutual inductance measurement of primary and secondary coils, and the Planck constant determination is described by a "joule balance" as
\begin{equation}
[M(z_2)-M(z_1)]I_1I_2-mg(z_2-z_1)=\int_{z_1}^{z_2}\Delta f(z)dz,
\end{equation}
where $M(z_1)$ and $M(z_2)$ denote the mutual inductances at positions $z_1$ and $z_2$; $I_1$ and $I_2$ are the currents in the primary and secondary coils. The advantage for joule balance is that the dynamic measurement is avoided, and hence the uncertainty arising from coil movement would be reduced. However, it requires a precision measurement of the mutual inductance.

A joule balance prototype has been developed at NIM for principle verification since 2006. In the prototype, a coil system was introduced for generating magnetic force and a conventional beam balance was applied in for weighing. A laser locking system including a piezoelectric crystal device was developed for the coil position control and a laser heterodyne interferometer was used to measure positions of the coil \cite{Yang2014}. In order to measure the DC value of the coil mutual inductance, two methods, the linear extrapolation at low frequency \cite{Li2011} and the standard square wave compensation method \cite{Lan2012}, have been developed. Two approaches can determine the mutual inductance with relative uncertainties of several parts in $10^7$ and agree within $1\times10^{-6}$.

\begin{figure}
\center
\includegraphics[width=3.1in]{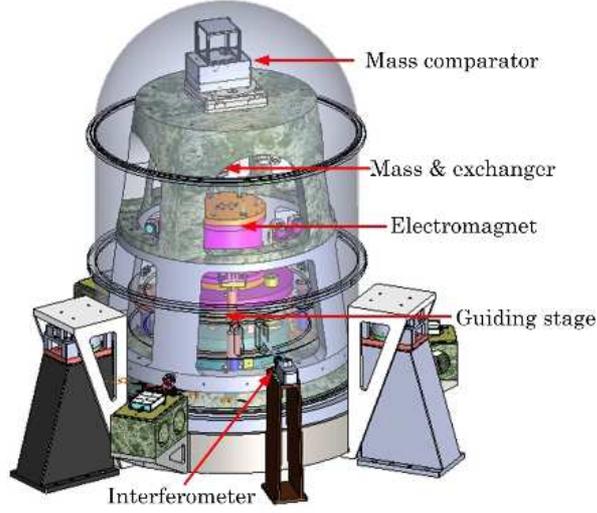}
\caption{\small New design of the NIM joule balance.}
\label{Fig.JB}
\end{figure}

The coil system is suitable for mutual inductance measurement, but a rather large DC current is required in order to generate enough magnetic force. As a result, a serious heating problem for the coil system has been observed, which became the main uncertainty source of the joule balance. In the beginning, a 250mA current was used to generate a 2N magnetic force, yielding more than 100W power consumption and 60ppm uncertainty. Later the coil system was optimized into a compact one to improve the utility. In this case, a 3N magnetic force was produced and the power of the coil was reduced to 40W, reducing the uncertainty to about 9ppm. The latest reported value of the Planck constant was $h=6.626104(59)\times10^{-34}$Js published in 2014 \cite{Zhang2014}.

Since 2013, a new joule balance platform (shown in figure \ref{Fig.JB}) has been under design and built. In the new system, a mass comparator will be used for weighing. The ferromagnetic coil system will be applied, which would increase the magnetic force to 5N and reduce the power assumption to about 9W. A vacuum system and a linear motor moving stage guiding the fixed coil are under development.

\subsection{MSL watt balance}
\begin{figure}
\center
\includegraphics[width=3.2in]{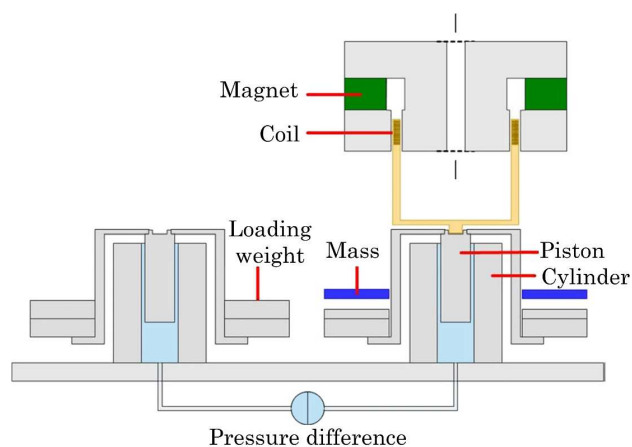}
\caption{\small Design of the MSL watt balance.}
\label{Fig.MSL}
\end{figure}

The MSL watt balance, shown in figure \ref{Fig.MSL}, is realized with several different ideas. Firstly, instead of a conventional weighing beam balance or commercial mass comparator, a twin pressure balance system was employed as the force comparator \cite{Sutton2009a}. In the design, the loaded piston can be freely rotating in a close-fitting vertical cylinder, and the pressure on the piston arising from the gas compression can be several newtons. The residual force was read from a differential pressure sensor in the weighing mode. Currently, the pressure balance has achieved a 5$\mu$g resolution and 15$\mu$g uncertainty for weighing 1kg test mass.

For most watt balances, in the velocity mode, they are operated either using a constant velocity or a quasi-constant velocity with constant induced voltage. In MSL watt balance, the alter moving of the coil was designed with low frequency, e.g., 0.1-10Hz \cite{Sutton2009}. The advantage for this approach is that the noise band can be obviously suppressed and the signal to noise ratio can be improved by using Fourier analysis. But in this condition, the amplitude of the induced voltage in coil is changing during the measurement, which may lead to difficulty in the voltage measurement, especially when the PJVS system is applied.

The third difference for MSL watt balance is on their magnet \cite{Sutton2014}. The magnet design is originated from the BIPM watt balance magnet with different permanent magnet locations. A ring-shaped permanent magnet and yoke arrangement is applied for generating uniform radial magnetic field in an annular gap. The design supplies higher flux concentration in the air gap and has some freedom in the choice of gap diameter. A disadvantage is that, even small gaps are added through inner and outer yokes, the systematic effect arising from the coil magnetic flux in the weighing mode should be considered carefully due to its relative low magnetic reluctance.

\subsection{KRISS watt balance}
\begin{figure}
\center
\includegraphics[width=2in]{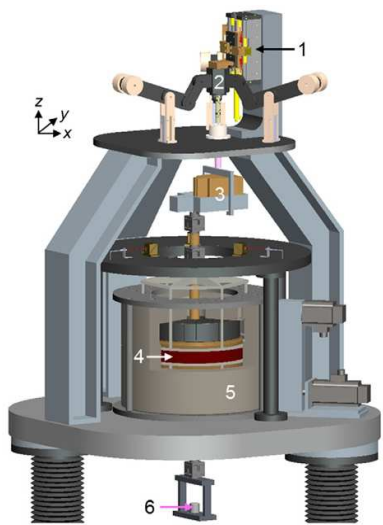}
\caption{\small Design of the KRISS watt balance. The numbers denote respectively 1-linear motor, 2-guiding stage, 3-weighing cell, 4-coil, 5-magnet and 6-test mass. Adopted from \cite{Kim2014}. }
\label{Fig.KRISS}
\end{figure}

After two-year planning, KRISS became a new member of the watt balance community in 2012. Based on combination and optimization of exciting watt balances techniques, the KRISS watt balance design is shown in figure \ref{Fig.KRISS} \cite{Kim2014}. The magnet is a typical two-permanent, one-coil structure, similar to the magnet of the BIPM watt balance group. The Ni-Fe alloy flux shunt is used to compensate the magnetic field change due to the temperature variations. A commercial weighing cell with capacity of 5 kg and resolution of 1$\mu$g from Mettler, Toledo is used in the weighing mode. A piston gauge moving stage is designed to guide the vertical motion the coil and weighing cell in the velocity mode, and a straightness of 1.2$\mu$m in $\pm$20mm has been achieved. The first run of the KRISS watt balance is expected in 2015.

\section{Progress of the Avogadro project}
As discussed in section 2.2, the Avogadro project measures the Avogadro constant $N_A$ and indirectly determines the Planck constant $h$ based on the precision measurement of their product $N_Ah$ \cite{CODATA2010}. Different from the watt balance experiment that can be carried out by an individual metrology institute, the Avogadro project is persuaded by an international collaboration between laboratories \cite{Andreas2011}, i.e., the International Avogadro Coordination (IAC). Related institutes include the Physikalisch-Technische Bundesanstalt (PTB, Germany), the National Metrology Institute of Japan (NMIJ, Japan), the METAS, the NIST, the Istituto Nazionale di Ricerca Metrologica (INRIM, Italy), the BIPM, the Institute for Reference Materials and Measurements (IRMM, Belgium), the NRC and the National Measurement Institute (NMIA, Australia). Each institute involves at least one of the key quantities for determining the Avogadro constant $N_A$.

\begin{figure}[h]
\center
\includegraphics[width=3.2in]{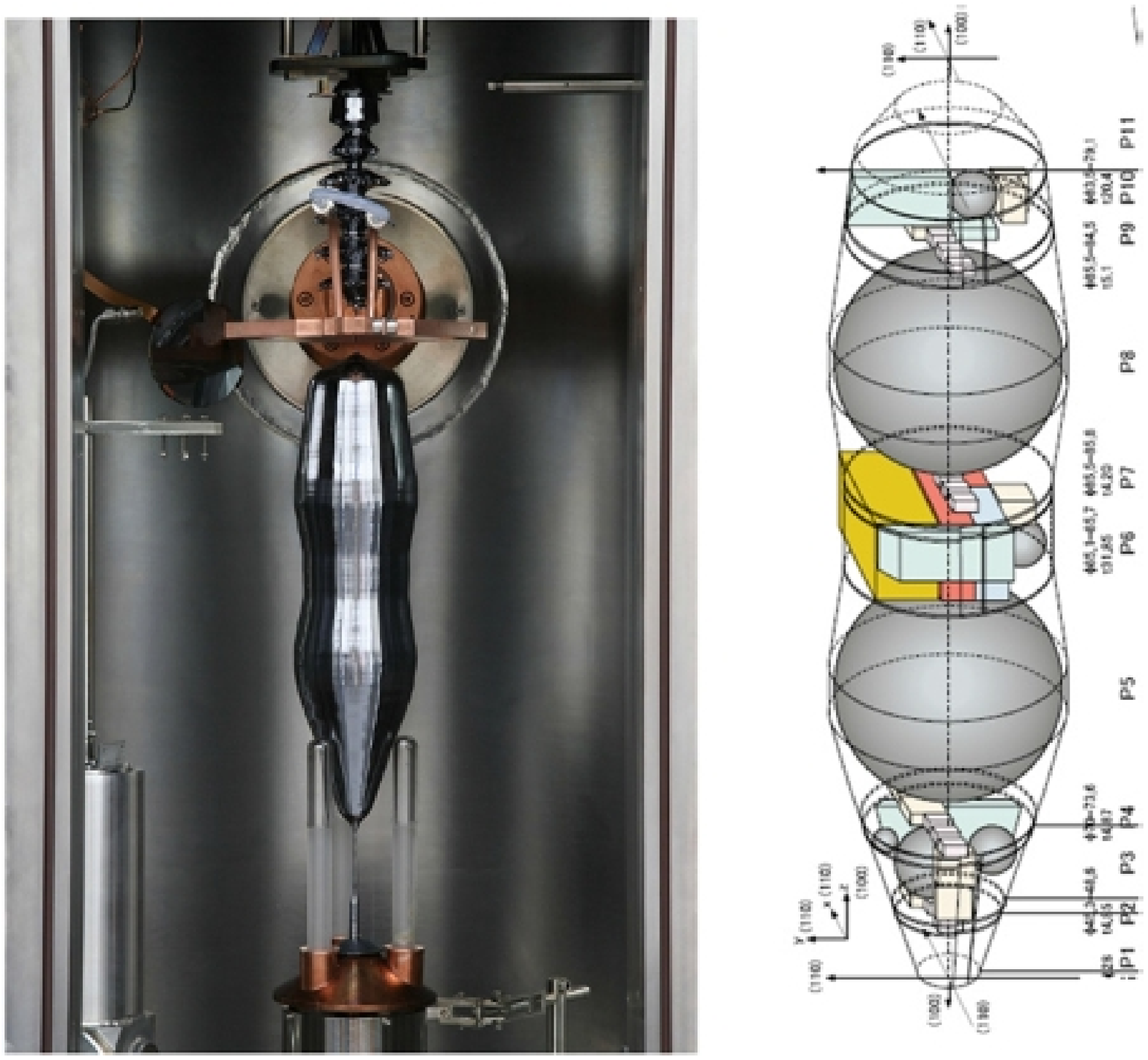}
\caption{\small The 5kg monocrystal of the enriched material (left) and the plan to be manufactured (right). Reproduced with permission \cite{Andreas2011}.}
\label{Fig.Na1}
\end{figure}

The first set measurement of the Avogadro project was based on three spheres made by the natural silicon crystal, which consists an isotopic distribution of 92.2\% $^{28}$Si, 4.7\% $^{29}$Si and 3.1\% $^{30}$Si respectively. The first measurement of the Planck constant $h=6.6260762(21)\times10^{-34}$Js with a relative uncertainty of $3.1\times10^{-7}$ was published by the natural silicon crystal in 2004 \cite{Becker2003Determination}. However, this result differs the value obtained from the watt balance experiment by about $1\times10^{-6}$. It was later found by new atomic spectrometer calibrations that the difference was contributed by the isotopic distribution of natural silicon and the value of the Planck constant was corrected to $h=6.6260681(16)\times10^{-34}$Js with an reduced uncertainty of $2.4\times10^{-7}$ \cite{Valkiers2011}.

For further reducing the uncertainty caused by the mole mass measurement, a second generation silicon spheres, in which the crystal was fabricated with 99.995\% $^{28}$Si, were developed \cite{becker2006}. The enriched $^{28}$Si material, supplied by the Institute of Chemistry of High-Purity Substances of the Russian Academy of Sciences (IChHPS-RAS) in Nizhny Novgorod and the Central Design Bureau of Machine Building (CDBMB) in Saint Petersburg, was grown into a 5 kg monocrystal (shown in figure \ref{Fig.Na1}) in the Leibniz-Institut f\"{u}r Kristallz\"{u}chtung (IKZ) in Berlin. Two pieces of the crystal numbered 5 and 8 were manufactured into two silicon spheres at the Australian Centre for Precision Optics (ACPO), named AVO28-S5 and AVO28-S8. The current measurement of the Avogadro project is mainly based on these two silicon spheres. Here the progress of the Avogadro project is reviewed by dividing the measurement into several aspects, i.e., measurements in mole mass, lattice parameter, mass and volume.

\subsection{Mole mass measurement}
The mole mass $M_0$ is determined following equation (\ref{eq.si}) by measurements of the crystal sample composition of three isotopic distributions, $^{28}$Si, $^{29}$Si and $^{30}$Si. Three methods have been developed for measuring the isotopic distribution. The gas mass spectrometry (GMS) of the SiF4 gas was applied at IRMM \cite{Mana2010} and the GMS of a direct fluorination by BrF5 was used at the Institute of Mineral Resources (IMR, China); the isotope dilution combined with multicollector inductively coupled plasma mass spectrometry (IDMS) was used in PTB \cite{Pramann2011}; the secondary ion mass spectrometer (SIMS) using a time-of-flight mass analyser was employed in the Institute for Physics of Microstructures of the Russian Academy of Sciences (IPM-RAC) \cite{Drozdov2010}.

As the total isotopic distribution of $^{29}$Si and $^{30}$Si is about 0.005\%, for achieving the measurement goal of $2\times10^{-8}$ for mole mass determination, the isotopic distribution should be measured with a relative uncertainty within $4\times10^{-4}$. Measurement results showed that the GMS method at IRMM and the IDMS method at PTB have achieved uncertainties less than $1\times10^{-8}$ for determining $M_0$. However, the measurement uncertainties do not overlap each other. The difference was found due to the chemical preparation of the samples \cite{Bulska2011}. By a functional analysis of adding natural silicon to the enriched silicon using IDMS method, three method would be consistent with each other. In the IAC determination, the data obtained by IDMS method at PTB was used. The molar mass values obtained were $27.97697017(16)$g/mol and $27.97697025(19)$g/mol for AVO28-S5 and AVO28-S8 respectively \cite{Andreas2011}.

For further investigation the difference of the mole mass determinations between PTB and IRMM, in 2011 NRC was invited
to perform an independent measurement of the isotopic composition of the enriched $^{28}$Si material. The mole mass was measured as $M_0=27.97696839(24)$g/mol at NRC \cite{Steele2012}, which obviously differed from the PTB determination. For  an arbitration, the mole mass of the enriched $^{28}$Si material has also been measured at NIST and NMIJ by the IDMS method. In 2014, both the NMIJ and NIST published the measurement results, $M_0(\mathrm{NMIJ})=27.97697009(14)$g/mol \cite{Narukawa2014} and $M_0({\mathrm{NIST}})=27.976969757(92)$g/mol \cite{Vocke2014}.
The NMIJ and NIST results are coconscious with the PTB result and confirm the IDMS method applied by PTB is capable of very high accuracy.

\subsection{Lattice parameter measurement}
In order to obtain the lattice parameter $a$ in equation (\ref{eq.Nh}), both the absolute and relative $d_{220}$ measurements have been carried out at NMIs. The first $d_{220}$ measurement was taken in 1973 at NIST \cite{Deslattes1973}, which was a milestone linking visible and X-ray wavelengths. Later, more accurate absolute $d_{220}$ values were obtained by the combined X-ray and optical interferometer (XROI) method at PTB in Germany \cite{Becker1981}, INRIM in Italy \cite{Becker2007}, and NMIJ in Japan \cite{Cavagnero2004}. As different samples of natural silicon crystals were used in NMIs, highly accurate measurements of the relative difference of $d_{220}$ were also taken at NIST \cite{Hanke2005} and PTB \cite{Becker2003}, whose measurement data were applied for corrections during international comparisons.

As the highly enriched $^{28}$Si material is now used in order to overcome difficulties in the molar mass measurement of a natural silicon crystal, the lattice parameter of the enriched $^{28}$Si has to be remeasured. In INRIM, the XROI method has been newly developed and the travel range of the crystal analyzer was increased to many centimeters \cite{Ferroglio2008}. Besides, a two-crystal Laue diffractometry has been developed in NIST for absolute $d_{220}$ measurement \cite{Massa2011S}. The absolute $d_{220}$ measurements at INRIM and NIST, which have relative uncertainties of $3.5\times10^{-9}$ and $6.5\times10^{-9}$ respectively, agreed each other within $2\times10^{-9}$ \cite{Massa2011}. Note that measurements showed the lattice parameter of the enriched silicon crystal was $1.9464(67)\times10^{-6}$ larger than that of the natural silicon crystal, and the reported experimental value has been found by E. Massa {\it et al} to be consistent with quantum-mechanics calculations \cite{Massa2011}.

In addition, the point defects should be measured and corrected during the determination of $d_{220}$. In reality, the impurity concentrations $N_i$ were measured by infrared spectroscopy \cite{Zakel2011}; the lattice expansion and contraction coefficients $\beta_i$, were measured by X-ray diffraction in significantly doped Si samples \cite{Becker2001}. It is found from the relative correction value $N_i\beta_i$ that the deviation from homogeneous lattice spacing was less than $1\times10^{-8}$.

\subsection{Mass measurement}
The mass of the silicon crystal is measured as part of determination of the silicon crystal density $\rho=m/V_m$. Note that $\rho$ in equation (\ref{eq.Nh}) should be the silicon crystal density excluding the oxide layer, hydrocarbon or other contaminations, point defects and water sorption. Therefore $m$ should be the sphere core mass calculated as
\begin{equation}
m=m_t-m_l+m_d,
\end{equation}
where $m_t$ is the total mass of the sphere; $m_l$ is the mass of the sphere surface layer; $m_d$ is the mass of point defects of the crystal.

The measurement of two $^{28}$Si sphere masses $m_t$, carried out by the BIPM, PTB and NMIJ, were compared to a 1kg platinum-iridium (Pt/Ir) mass standard. The masses of the $^{28}$Si spheres determined in air and under vacuum showed a good agreement in the three involved laboratories \cite{Picard2011}, and a combined relative uncertainty of $5\times10^{-9}$ was achieved.

The surface layer of the silicon spheres was modeled into four sub-layers \cite{Busch2011}, named the carbonaceous contamination layer (CL), the chemisorbed water layer (CWL), the layer of metallic silicides (ML) and the pure oxide layer (OL) respectively. The mass, thickness and chemical composition of the surface layer were characterized several different methods \cite{Andreas2011}, including the X-ray fluorescence (XRF) analysis, the near-edge X-ray absorption fine structure (NEXAFS) spectroscopy, the X-ray reflectometry (XRR), the X-ray photoelectron spectroscopy (XPS) and the optical spectral ellipsometry (SE) \cite{Zhang2012}. The CL thickness was measured by XRF with a reference sample having a 6.5nm carbon
layer \cite{Seah2004}, and the CL mass obtained was about 14$\mu$g with 0.5nm thickness. The CWL was determined by gravimetric measurement with 7.7$\mu$g and 0.28nm thickness \cite{Mizushima2004}. The ML mass 107.5$\mu$g with 0.54nm thickness was characterized by XRF, NEXAFS and XPS \cite{Busch2011}. The  OL was measured about 90$\mu$g with 1.4nm thickness by XPS \cite{Busch2009}. The combined masses of surface layer $m_l$ were determined as 222.1(14.5)$\mu$g and 213.6(14.4)$\mu$g for AVO28-S5 and AVO28-S8 respectively.

The mass correction $m_d$ arising from the point defects is determined as
\begin{equation}
m_d=V_m\sum(m_{28}-m_i)N_i,
\label{eq.md}
\end{equation}
where $m_{28}$ and $m_i$ are the masses of the $^{28}$Si atom and of the $i$th point defect. Following equation (\ref{eq.md}), $m_d$ corrections of 8.1(2.4)$\mu$g and 24.3(3.3) $\mu$g were applied for the AVO28-S5 and the AVO28-S8.

\subsection{Volume measurement}

\begin{figure}[h]
\center
\includegraphics[width=2.9in]{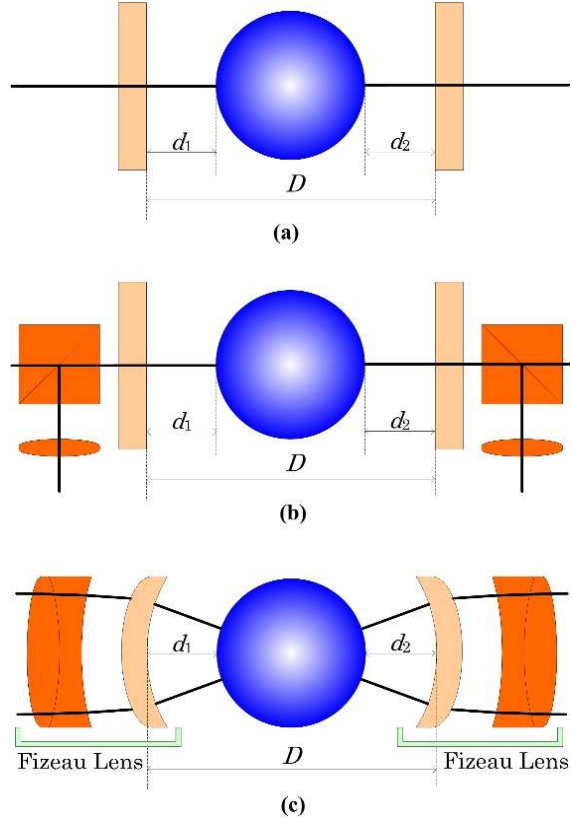}
\caption{\small Interferometers to measure the diameter of spheres. (a) Optical configuration in Saunders' interferometer, (b) optical configuration in the NMIJ interferometer and (c) optical configuration in the PTB interferometer. }
\label{Fig.d}
\end{figure}

The first measurement for precision determination of the sphere volume was taken at NIST by an optical interferometer which was designed by Saunders and shown in figure \ref{Fig.d}(a) \cite{Saunders1972}. Later a two-optical-interferometer system was developed at INRIM \cite{Sacconi1989} and NMIJ \cite{Fujii1992} as shown in figure \ref{Fig.d}(b), in which the reflected light from the sphere was collimated by a lens to overcome the problem due to the diffraction of the reflected light in Saunders' interferometer \cite{Masui1997}. At PTB, a Fizeau interferometer with a spherical etalon was developed shown in figure \ref{Fig.d}(c) \cite{Bartl2009} with an advantage in analysing the entire surface of the sphere \cite{ZhangJT2011}.

The measurement of the sphere diameter was differential and can be divided into two steps: 1) measuring the etalon dimension $D$ without a sphere and 2) measuring two gaps $d_1$ and $d_2$ of a sphere and the relevant plate. The sphere diameter is determined as
\begin{equation}
d=D-d_1-d_2.
\end{equation}

\begin{figure}
\center
\includegraphics[width=3.5in]{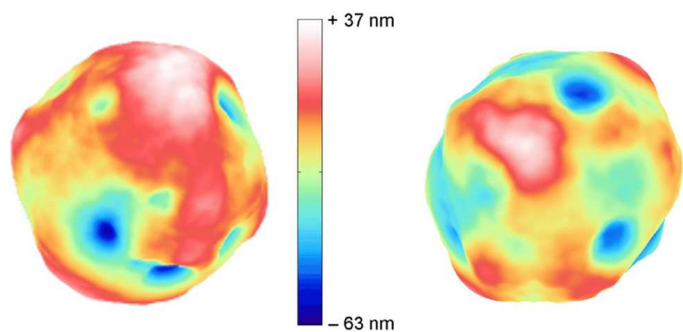}
\caption{\small Diameter topographies of two enriched $^{28}$Si spheres. The left and right denote the AVO28-S5 and AVO28-S8 respectively. Reproduced with permission \cite{Andreas2011}.}
\label{Fig.NAV}
\end{figure}

The sphere diameter measurement has been carried out in NMIA \cite{Kenny2001}, NMIJ \cite{Kuramoto2011}, and PTB \cite{Bartl2011}, respectively employing the methods shown in figure \ref{Fig.d}(a), \ref{Fig.d}(b), and \ref{Fig.d}(c). At NMIJ,  single diameter values were obtained by determination of the central order of the concentric fringes. And in PTB, the interference patterns equal to the thickness fringes in a full view, which allows a complete topographical mapping of both spheres shown in figure \ref{Fig.NAV}. The comparison between NMIJ and PTB showed excellent agreement of several nanometers when $d\approx93722.9720\mu$m.

\subsection{Results}
In 2011, the Avogadro project published the first measurement result IAC-2011 by enriched silicon spheres \cite{Andreas2011}. The Planck constant obtained is $h=6.62607014(20)\times10^{-34}$Js with a relative uncertainty of $3.0\times10^{-8}$. Note that this value is between the latest results produced by the watt balance experiment, which is $3.0\times10^{-8}$ lower than the NRC-2014 watt balance measurement and $5.3\times10^{-8}$ higher than the NIST-2014 watt balance result.

By performing an independent measurement of the isotopic composition of the enriched $^{28}$Si material, when NRC produced its initial watt balance result in 2012, a determination of the Planck constant by enriched silicon sphere, $h=6.62607055(21)\times10^{-34}$Js with relative uncertainty of $3.0\times10^{-8}$ was also published \cite{Steele2012}. This value is consistent with the NRC watt balance result but $6.2\times10^{-8}$ higher than the IAC-2011 result.

For further checking the difference between IAC and NRC determination, the mole mass of the enriched $^{28}$Si material has also been measured at NIST and NMIJ. In 2014, NIST published its determination of the Planck constant $h=6.62607017(21)\times10^{-34}$Js \cite{Vocke2014}, which is consistent with the NMIJ result $h=6.62607011(22)\times10^{-34}$Js reported in the same year \cite{Narukawa2014}.
Both the NIST and NMIJ results agreed well with the IAC-2011 result.

There is a double that the measurement result of the Avogadro project may dependent on the enriched $^{28}$Si material. For further investigation of this effect, a new round of measurement is being carried out. PTB has purchased 2 different single $^{28}$Si crystals of 5kg each  from Russia, and four silicon spheres are being under construction. It is expected in end of 2014, the first two silicon spheres would be ready for measurements.

\section{Conclusion and discussion}
\subsection{Summary of the Planck constant determination}
\begin{figure}[h]
\center
\includegraphics[width=2.6in]{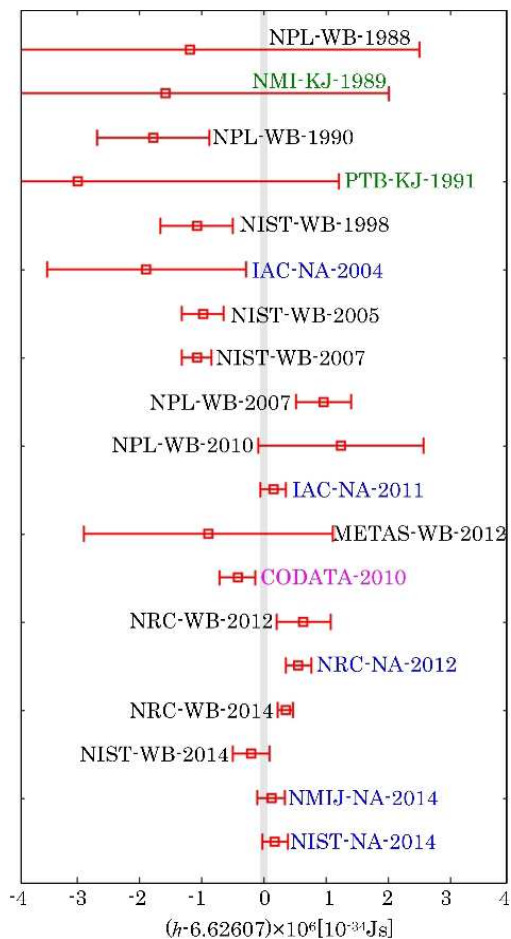}
\caption{\small Summary of numeral determinations of the Planck constant with relative uncertainties smaller than $1\times 10^{-6}$. KJ, WB, and NA present the result obtained from the voltage balance, the watt balance, and the Avogadro project. }
\label{Fig.h}
\end{figure}

Figure \ref{Fig.h} lists the numeral determinations of the Planck constant with relative uncertainties smaller than $1\times 10^{-6}$ \cite{CODATA2006, CODATA2010, Steele2012, Sanchez2014, Schlamminger2014, Narukawa2014, Vocke2014}. According to CIPM 2013 recommendations, the numeral value of the Planck constant will be fixed only when four conditions of the approach as follows have been achieved.
(1) At least three independent experiments, including work from watt balance and
XRCD experiments, yield consistent values of the Planck constant with relative
standard uncertainties not larger than 5 parts in $10^8$;
(2) At least one of these results should have a relative standard uncertainty not larger
than 2 parts in $10^8$;
(3) The BIPM prototypes, the BIPM ensemble of reference mass standards, and the
mass standards used in the watt balance and XRCD experiments have been
compared as directly as possible with the international prototype of the kilogram;
(4) The procedures for the future realization and dissemination of the kilogram, as
described in the mise en pratique, have been validated in accordance with the
principles of the CIPM Mutual Recognition Arrangement (MRA).


It is concluded from figure \ref{Fig.h} that the NRC-WB-2014 result has achieved the CIPM recommendation (2) with relative uncertainty of $1.9\times10^{-8}$; the NRC-WB-2014 result, the NIST-WB-2014 result and the IAC-NA-2011 result have achieved the CIPM recommendation (1) with relative uncertainty of $1.9\times10^{-8}$, $4.5\times10^{-8}$, and $3.0\times10^{-8}$ respectively. The mean weighted value for the Planck constant $h=6.626070235(97)\times10^{-34}$Js with relative uncertainty of $1.5\times10^{-8}$ is calculated based on the NRC-WB-2014 result, the NIST-WB-2014 result and the IAC-NA-2011 result.
Recommendation (3) is being carried out and (4) is under planing, and their results are expected to be approved in 2015 and 2017 respectively.

 Although recommendations (1) and (2) have been met, the total set of data is still inconsistent.
The difference between these results is caused by certain systematic errors. During the latest years, some efforts have been paid on investigating possible systematic errors of both a watt balance and the Avogadro project, e.g., \cite{shisong2013, shisong2014NIST, Mana2013, Mana2014}. In future, on the one hand, further and detailed investigation on current operated watt balance and Avogadro project should continue; on the other hand, other methods for precision measurement of the Planck constant should be encouraged \cite{Kibble2014}.


\subsection{Towards a revised physical constant system}
The new kilogram definition based on fixing the numeral value of the Planck constant will eliminate both time and space dependencies of an artifact standard, and hence the long-term stability of the whole SI fundamentals is improved. After the Planck constant is fixed (exact number with zero uncertainty), the uncertainties of some other constants, which are related to the Planck constant, will synchronously change following basic physical laws. As a result, this transition will lead to a revised physical constant
system also. Table \ref{constant} summarizes the uncertainty change of some physical constant that are related to the Planck constant \cite{Newell2014R}.
\begin{table}[h]
\small
\caption{\small The uncertainty change of some physical constants that are related to the Planck constant in the revised SI. $u_c$ and $u_n$ denote the relative uncertainty in the current SI and the revised SI respectively.}
\label{constant}
\centering
\begin{tabular}{cccc}
\hline
physical constant & symbol & $u_c$($\times10^{-8}$) & $u_n$($\times10^{-8}$)\\
\hline
the electron charge & $e$ & 2.2 & 0\\
the Avogadro constant & $N_{A}$ & 4.4 & 0\\
the von Klitzing constant& $K_{{J}}$ & 2.2 & 0\\
the Josephson constant& $R_{{K}}$ &0.032 & 0\\
the Faraday constant& $F$ & 2.2 & 0\\
the electron mass& $m_{e}$ & 4.4 & 0.064\\
the unified atomic mass unit& $m_{\mu}$ & 4.4 & 0.07\\
the Bohr magneton&$\mu_{{B}}$&2.2& 0.064\\
the nuclear magneton&$\mu_{{N}}$ &2.2&0.07\\
the fine structure constant& $\alpha$ & 0.032 & 0.032\\
the magnetic constant& $\mu_{0}$ & 0 & 0.032\\
the vacuum permittivity& $\varepsilon_{0}$ & 0 & 0.032\\
the impedance of free space& $Z_{0}$ & 0 & 0.032\\
\hline
\end{tabular}
\end{table}

It can be seen from Table \ref{constant} that except for a small uncertainty increase for the magnetic constant $\mu_0$, the vacuum permittivity $\varepsilon_{0}$ and the impedance of free space $Z_0$, the uncertainty for all the other physical constant listed has been greatly reduced. Note that during the transition of a revised physical constant system, the uncertainty of the fine structure constant $\alpha$ will not change. It is also notable that two typical uncertainty components after redefinition, $6.4\times10^{-10}$ (also most part of $7\times10^{-10}$) and $3.2\times10^{-10}$, are observed, which are mainly contributed by the fine structure constant, i.e., $2u(\alpha)$ and $u(\alpha)$. Therefore, it is predicted that after the revision, the accurate measurement of the fine structure constant would be a next task towards a more precision physical constant.

\subsection{Outlooks}
Both the watt balance and the Avogadro project are ongoing. According to the CIPM roadmap towards a redefinition in 2018 \cite{Milton2014}, more numeral determinations of the Planck constant at NMIs will be produce in the near future. Based on the current $h$ measurement results, it is dare to say that a time for determining the Planck constant at $10^{-8}$ level has come. For watt balance, further investigations on possible systematic effects among different NMIs should in more details to reduce both their biases and uncertainties. For the Avogadro project, a new round of measurement for the enriched $^{28}$Si should be taken to prove the data stability.

\section*{References}


\end{document}